\documentclass[12pt]{article}%
\usepackage{amsmath,amssymb,amsthm,amsfonts}
\usepackage{wasysym}
\usepackage{graphicx}
\usepackage[dvipsnames]{xcolor}
\usepackage{stackengine}

\usepackage[colorlinks]{hyperref}
\usepackage{tikz}
\usepackage{subcaption}
\usepackage{multirow}

\usepackage{appendix}

\newcommand{\ea}{\textit{et al. }}
\renewcommand{\epsilon}{\varepsilon}
\renewcommand{\th}{\text{th}}

\definecolor{red}{rgb}{0.8500, 0.3250, 0.0980}
\definecolor{green}{rgb}{0.4660, 0.6740, 0.1880}
\definecolor{yellow}{rgb}{0.9290, 0.6940, 0.1250}
\definecolor{blue}{rgb}{0, 0.4470, 0.7410}

\begin{document}

\title{Damped-driven system of bouncing droplets leading to deterministic diffusive behavior}

\author{Aminur Rahman \thanks{Corresponding Author, \url{arahman2@uw.edu}},
\thanks{Department of Applied Mathematics, University of Washington}}
\date{}

\maketitle

\begin{abstract}
Damped-driven systems are ubiquitous in science, however the damping and driving mechanisms are often quite convoluted.  This manuscript presents an experimental and theoretical investigation of a fluidic droplet on a vertically vibrating fluid bath as a damped-driven system.  We study a fluidic droplet in an annular cavity with the fluid bath forced above the Faraday wave threshold.  We model the droplet as a kinematic point particle in air and as inelastic collisions during impact with the bath.  In both experiments and the model the droplet is observed to chaotically change velocity with a Gaussian distribution.  Finally, the statistical distributions from experiments and theory are analyzed.  Incredibly, this simple deterministic interaction of damping and driving of the droplet leads to more complex Brownian-like and Levy-like behavior.
\end{abstract}

\section{Introduction}
\label{Sec: Intro}

Damped-driven systems in physical and biological sciences often lead to very complex phenomena, which nevertheless can be modeled as mathematically tractable deterministic dynamical systems \cite{Duffing1918, vanderpol1927, HodgkinHuxley1952, Lorenz1963, Holmes82, MCT84, kutz2006mode, koch2020mode}.  It has also been shown that the interplay between damping and driving can induce diffusive behavior such as in the driven Josephson junction \cite{HubermanJosephsonJunction1980, ReithmayerThesis}.  Interestingly, the models for the driven Josephson junction \cite{HubermanJosephsonJunction1980, ReithmayerThesis} are deterministic, whereas Einstein derived a theory for Brownian motion as a stochastic process \cite{EinsteinDiffusion}.  Other examples of diffusive behavior induced through chaotic dynamical systems are treated in reviews by Geisel \cite{DeterministicDiffusionReview} and by Artuso and Cvitanovi\'{c} \cite{ChaosBookDeterministicDiffusion}.  While all chaotic systems are seemingly random \cite{Strogatz94}, deterministically diffusive systems experience Brownian-like behavior \cite{DeterministicDiffusionReview} where the mean squared displacement of the diffusive variable increases monotonically with time and admits a Gaussian distribution (during normal diffusion) for the variable from the initial point.

Since the seminal work of Couder and coworkers \cite{CPFB05, CouderFort06}, bouncing droplets have demonstrated fundamental physical phenomena at the macro-scale as detailed in the reviews by Bush and coworkers \cite{Bush15a, Bush15b, BushOza20_ROPP}.  In this system we drop a silicon oil droplet onto a silicon oil bath sitting on top of a vertically vibrating shaker.  It has been known since the 1970s that such a droplet can bounce on the bath without coalescing \cite{Walker1978}.  If the force on the shaker is increased the droplet starts moving horizontally (and is thus dubbed a \emph{walking droplet}) \cite{CPFB05, CouderFort06}.  In more recent years these bouncing droplets have shown quite complex self-organizing behavior such as forming crystal-like structures  \cite{EddiCrystal2009, EddiCrystal2011, ThomsonCrystal2020}.  If the forcing is increased furthermore still, Faraday waves form on the surface of the bath \cite{faraday_peculiar_1831, MWWAK_FaradayWaves1997, WWPMK01, ESMFRC11, TambascoDiffusion}, upon which the droplet can bounce around chaotically.

While the detailed hydrodynamic models can often be quite complex, reduced dynamical systems models provide a framework to analytically study the solution space of droplet trajectories thereby facilitating rigorous analysis of long-time statistics of the trajectories.  Stroboscopic models \cite{ORB13, OWHRB14, OHRB14} were some of the first model reductions allowing rigorous dynamical systems analysis of long-time statistics.  These models assume a closed form formula for the eigenmodes that contribute to the wavefield, and a continuous horizontal forcing on the droplet based on this wavefield.  Furthermore reductions came from discrete dynamical models \cite{Gilet14, RahmanBlackmore16, Gilet16, Rahman18, RahmanKutzDampedDriven} and discrete time differential equations \cite{DureyMilewski2017}.  Gilet's model of a walker in a linear confined geometry \cite{Gilet14} considered discrete impacts with a simplified wavefield due to the confined eigenmodes.  The model also only kept track of the droplet position at each impact and not the motion between impacts.  The simplicity of the model allowed for detailed dynamical systems analysis of the bifurcations and chaotic strange attractors \cite{RahmanBlackmore16, RJB17, RahmanBlackmore20}, and provided insight for the potential route to chaos.  Gilet later developed a discrete dynamical model of walkers on a circular corral \cite{Gilet16}, which produced long-time statistics similar to that of experiments \cite{HMFCB13}, but was more computationally efficient than the stroboscopic model.  In \cite{Rahman18} Rahman developed a standard map-like model for walkers on an annulus, which used the experiments of Filoux \ea \cite{FHV15Arxiv, FHSV17} as inspiration.  Through this 1-dimensional model he showed that the walker becomes chaotic through a cascade of period doubling bifurcations, which had previously been observed and hypothesized in experiments \cite{WMHB13}.  A more detailed overview of the dynamical systems models and techniques used to study walking droplets can be found in the review of Rahman and Blackmore \cite{RahmanBlackmoreReview20}.

Recently, researchers have observed potentially diffusive behavior in systems of walking droplets.  Tambasco \textit{et al.} experimentally studied the dynamics of a droplet in 2-dimensional free space above the Faraday wave threshold, and calculated an effective diffusivity for a droplet during erratic walking \cite{TambascoDiffusion}.  Later, Hubert \textit{et al.} developed an experiment-inspired stroboscope-like model for walkers below the Faraday wave threshold and studied how the effective diffusivity changes with the memory of the system \cite{HubertDiffusion, HubertDiffusion2022}.  In more theoretical studies, Gilet observed diffusive behavior in discrete dynamical models below the Faraday wave threshold \cite{Gilet14, Gilet16} and Durey observed such behavior in Lorenz-type continuous dynamical systems \cite{Durey2020}.  Moreover, Durey \textit{et al.} observed seemingly random walks in a stroboscopic model for speed oscillations \cite{DTB20} and Valani \textit{et al.} observe diffusive behavior in the timeseries of their stroboscopic model for a hypothetical classical wave-particle entity (akin to walkers) \cite{Valani2021}. Diffusive behavior has also been studied in several other works on walkers \cite{Bacot_PRL2019_Diffusion, DMB2018, DureyBush2021, Devauchelle_Diffusion2020, TCB18}.  Heretofore, diffusive behavior had not been systematically studied for a bouncer.   In this manuscript, we study experiments with bouncing droplets in an annulus above the Faraday wave threshold, and construct a hydrodynamic-kinematic model incorporating the evolution of the wavefield and both the horizontal and vertical dynamics of the droplet.

The remainder of the manuscript is organized as follows:  Section \ref{Sec: Kinematics} starts us off with a first principles model of the hydrodynamic-kinematic interactions of the droplet and the bath.  We discuss the numerics in Sec. \ref{Sec: Numerics} and the theoretical results for the hydrodynamic-kinematic model in Sec. \ref{Sec: Theoretical results}.  We conclude with some final discussions in Sec. \ref{Sec: Conclusion}.

\section{Experimental Setup}
\label{Sec: Experiments}

\begin{figure}[htbp]
\centering
\includegraphics[width=0.9\textwidth]{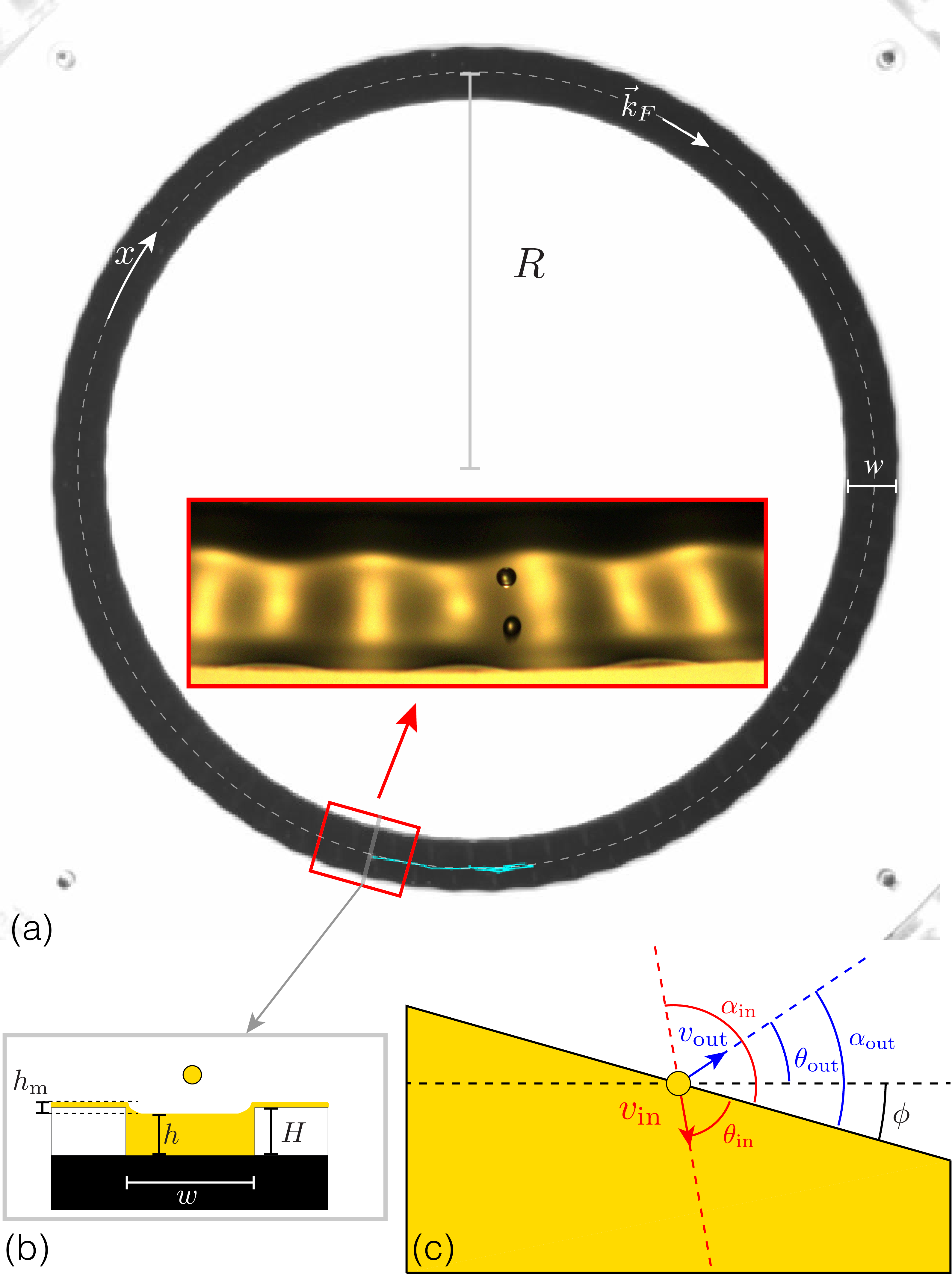}
\caption{A droplet bouncing on a quasi-1D standing wave. (a) Top view of the liquid bath with the annular channel (in black). The droplet recent trajectory is in cyan. Red frame: snapshot of a droplet with diameter $0.56$~mm while it is bouncing on the standing wave with $\gamma/\gamma_F=1.007$. (b) Vertical section of the liquid bath. (c) Schematics of the droplet bouncing for the theoretical model.}
\label{fig1}
\end{figure}

The schematic for the experimental set-up of Pucci \cite{GiuseppePrivate}, conducted at the MIT Applied Math Laboratory and privately communicated to the author by Pucci himself, are presented in Fig. \ref{fig1}(a,b). A circular container is filled with silicone oil with density $\rho=950$ kg/m$^3$, viscosity $\nu=20.9$ cSt and surface tension $\sigma=20.6$ mN/m.  The liquid bath has overall diameter $15.8$ cm and is divided in three regions with different liquid depths: an inner and an outer region (in white in Fig. \ref{fig1}(a)) in which only a thin liquid layer is present, and an annular channel (in black in Fig. \ref{fig1}(a)). The radius of the channel's centerline is $R = 6.63$~cm, the channel's width is $w = 7.5$~mm and its height $H = 4.6$~mm. The container is initially filled with silicone oil up to a height $\gtrsim H$. Then some liquid is removed from the channel until the liquid level in the channel reaches the height $h = 4.3$~mm. With this procedure, we create a tiny liquid meniscus with height $h_m \approx 0.3$~mm that will impede the drop to escape the channel, while the rest of the bath remains covered by a thin layer of oil (Fig. \ref{fig1}(b)). Liquid heights are measured with a dipping tip controlled by a micrometer with measurement error $\Delta h = 0.03$ mm.

The bath is vertically driven by an electromagnetic shaker with acceleration $\Gamma(t)=\gamma \cos(2 \pi f_0 t)$, where the vibrational frequency is $f_0=80$ Hz. The vibration system ensures constant acceleration amplitude to within $\pm 0.002$ g and spatially uniform vertical vibration to within 0.1$\%$ \cite{harris2015generating}. The vibration is monitored by two accelerometers placed symmetrically with respect to the center of the container. The experimental uncertainty on the dimensionless acceleration $\gamma/\gamma_F$ is less than 0.2\%, and primarily due to small variations of $\gamma_F$ during the course of the experiments, as result from slow drifts in bath temperature and the dependence of fluid viscosity on temperature. For $\gamma \geq \gamma_F \approx 4.2$~g, a quasi-1D standing surface wave with wavelength $\lambda_F = 4.75$~mm, wave vector $\vec{k}_F$ tangential to the channel centerline and period $T_F=2/f_0$ appears in the annular channel. This is a Faraday wave, that is, a manifestation of the Faraday instability \cite{faraday_peculiar_1831}, and is consistent with the dispersion relation of gravity-capillary waves \cite{guyon2015physical}.

A droplet of the same silicone oil with diameter $0.56$~mm is generated via a droplet generator \cite{harris_low-cost_2015} and placed in the channel while $\gamma < \gamma_F$. The container is sealed with a transparent acrylic lid that provides isolation from ambient air currents, a necessary precaution for repeatable experiments \cite{pucci_walking_2018}. The forcing acceleration is then increased above $\gamma_F$ and the droplet starts moving along the channel. The droplet remains at all time in the vicinity of the channel centerline, thus providing a good approximation for one-dimensional motion (Fig. \ref{fig1}(a)).  The droplet motion is recorded from above with a CCD camera at 20 frames per second and tracked with an in-house algorithm.
%---%

%---FIGURE---%
%---%

\section{Hydrodynamic-kinematic model}\label{Sec: Kinematics}

We model the experiment in Sec. \ref{Sec: Experiments}, a walking droplet on an annulus bouncing on a vibrating fluid bath, as a solid particle interacting inelastically with the waves created by the bath, and subsequently traveling through the air as a projectile.  Since the Reynolds number ($Re$) for the droplet during flight will be in the intermediate regime ($1 < Re < 10^3$), there are some nontrivial considerations for the drag on the droplet, which Molacek and Bush \cite{MolBush13a, MolBush13b} analyzed extensively.  Through their analysis it is evident that the correction terms to Stoke's drag is sufficiently small, and does not alter the qualitative behavior of the projectile.  Next, we assume the walker comes in contact with the wave at isolated impacts $n$ at a position $(x_n,y_n)$; that is, there is a single impact in a temporal neighborhood of the impact $t \in (t_n - \varepsilon, t_n + \varepsilon)$ for some $\varepsilon > 0$.  For the intermediate dynamics (the motion between impacts), let $\xi$ be the circumferential direction along the annulus and $\eta$ be the height, then our model becomes
\begin{subequations}
\begin{align}
\ddot{\xi} &= -\nu\dot{\xi};\qquad \xi(0)=x_n,\quad \dot{\xi}(0)=v_n^x\\
\ddot{\eta} &= -g - \nu\dot{\eta};\qquad \eta(0) = y_n,\quad \dot{\eta}(0) = v_n^y,
\end{align}
\label{Eq: ProjectileODEs}
\end{subequations}
where $\nu = 6\pi\mu_a/(\rho R_d^2)$ is the coefficient due to Stoke's drag for a sphere of radius $R_d$ with a material density of $\rho$ through air with dynamic viscosity $\mu_a$.  furthermore, $v_n^x$ and $v_n^y$ are the velocities, in the $x$ and $y$ directions respectively, immediately after the $n^\th$ impact.

\begin{figure}[htbp]
\centering
\stackinset{l}{1mm}{t}{1mm}{\textbf{ (a)}}{\stackinset{l}{-4mm}{t}{1.45in}{\textbf{ $\eta$}}{\stackinset{l}{0.78in}{b}{-4mm}{\textbf{ $\xi$}}{\includegraphics[height = 0.6\textheight]{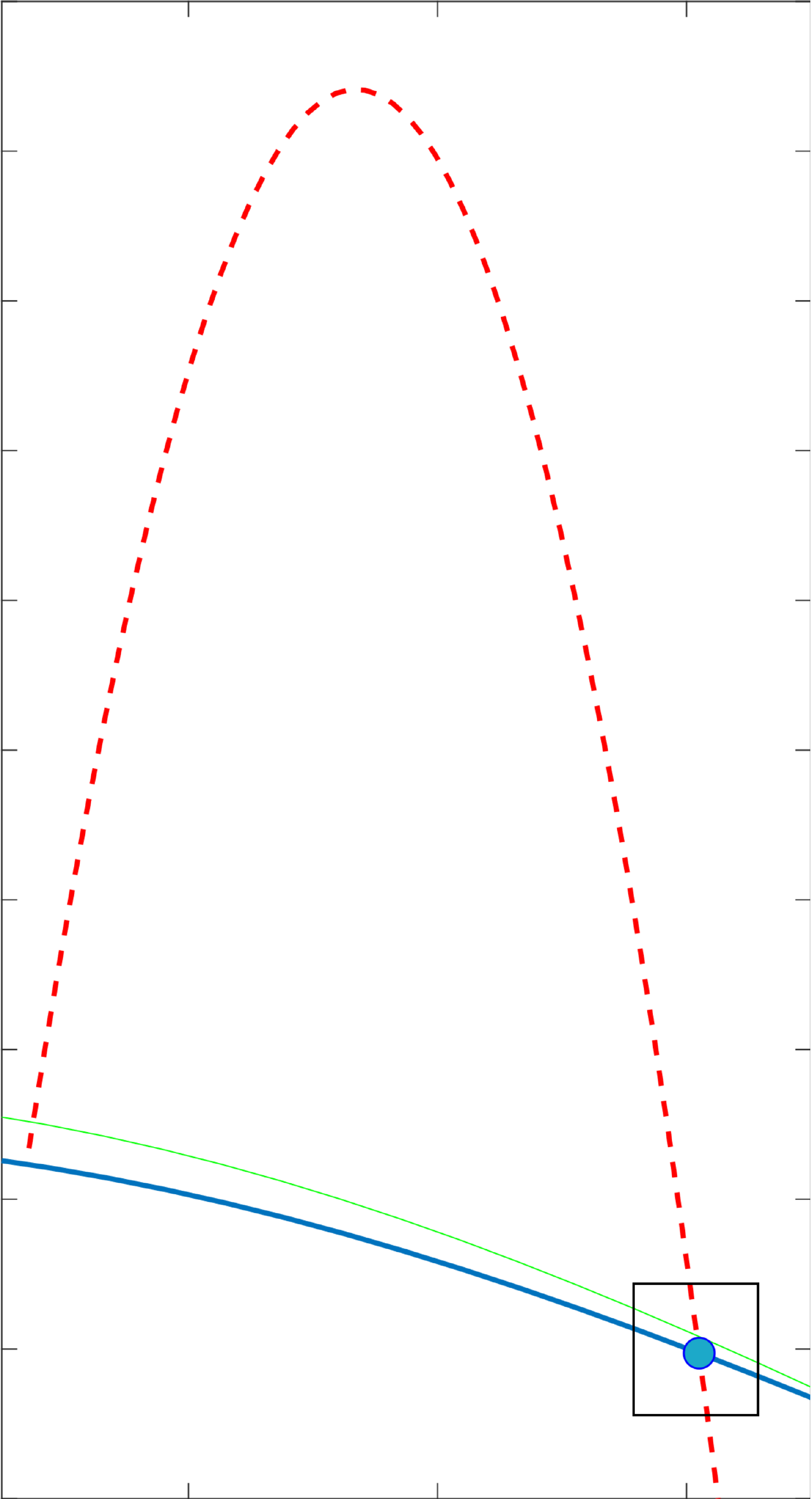}}}}
\refstepcounter{subfigure}\label{Fig: Path}
\stackinset{l}{}{t}{}{\textbf{ (b)}}{\includegraphics[height = 0.6\textheight]{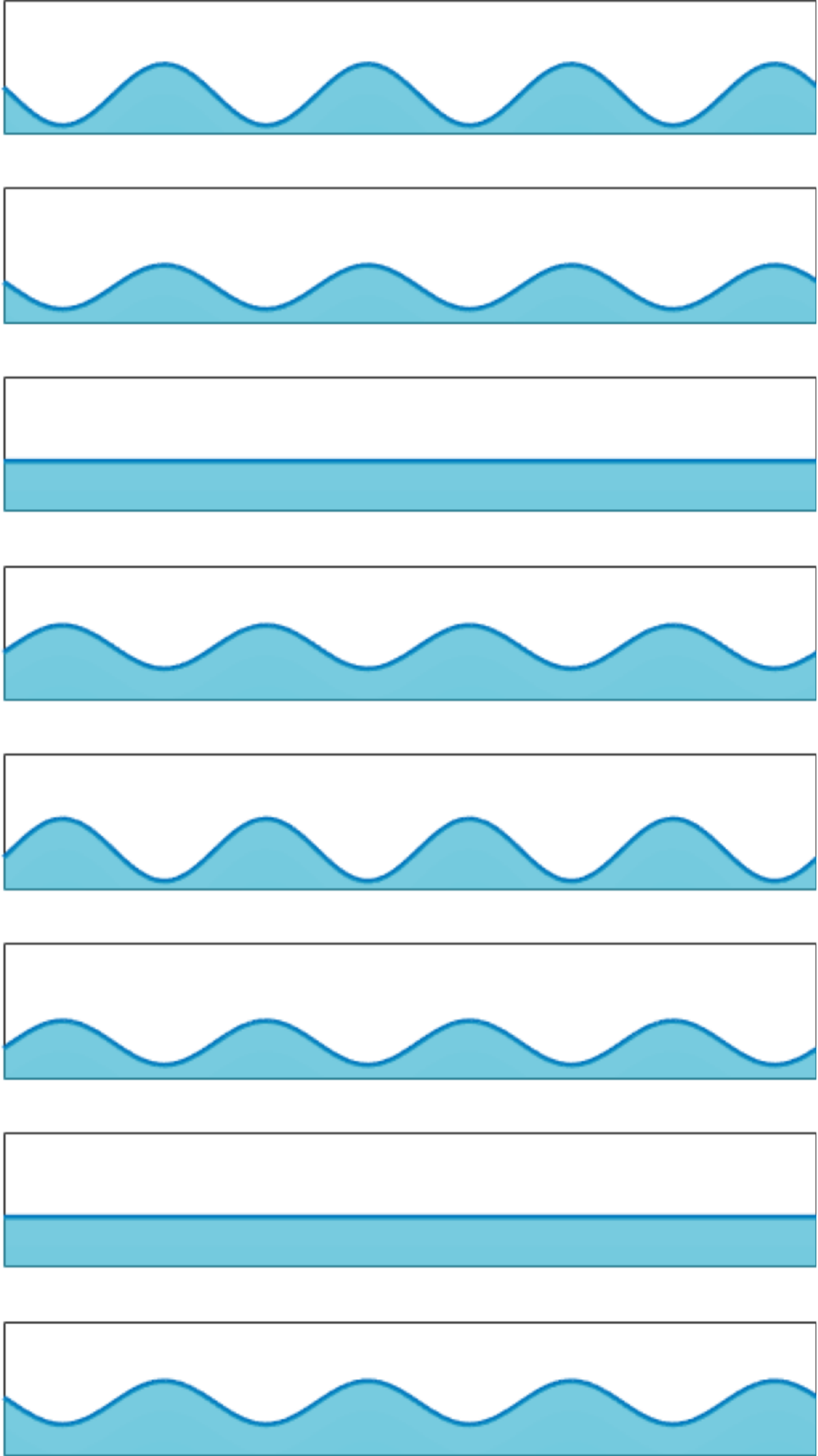}}
\refstepcounter{subfigure}\label{Fig: Waves}
\caption{Numerically computed flight path of the walker and evolution of the wavefield.
\textbf{(a)} The flight path (red dashed curve) of the walker (solid blue circle) from the $(n-1)^\th$ impact, shown as the intersection between the path and the wavefield at the time of the $(n-1)^\th$ impact (solid green upper curve), to the $n^\th$ impact, shown as the intersection between the path and the wavefield at the time of the $n^\th$ impact (solid blue lower curve).  The box around the walker indicates the region that will be magnified for analysis in Fig. \ref{Fig: Angles}.  \textbf{(b)} The evolution of the wavefield over one cycle from time $t = 0$ to $t = 7/4f$ at intervals $t = 1/4f$.}\label{Fig: Paths and Waves}
\end{figure}

Since the intermediate dynamics only considers the time between impacts, we use the time $t$ for the time between impacts $n$ and $n+1$, and $T = \sum_{i=0}^n t_i+ \Delta t;\qquad \Delta t\in (t_n,t_{n+1})$ as the elapsed time for the evolution of the wavefield.  In the experiments we observe that the wavefield, illustrated in Fig. \hyperref[Fig: Waves]{\ref{Fig: Paths and Waves}b}, is of the form
\begin{equation}
\begin{split}
\Psi(T,x) = &A\cos(\pi fT)\cos\left(\frac{2\pi}{\lambda_f}x\right)\\
&+\frac{\gamma}{(2\pi f)^2}\cos\left(2\pi f[T-\varphi]\right).
\end{split}
\label{Eq: Wave}
\end{equation}
where $\gamma$ is the maximum bath acceleration, $f$ is the bath frequency, which implies a period of $T_f = 2/f$, $\varphi$ is the bath vibration phase, and $\lambda_f$ is the wavelength.

Next, we nondimensionalize \eqref{Eq: ProjectileODEs} using $\tau = ft/2 = t/T_f$, $\hat{\xi} = \xi/\lambda_f$, and $\hat{\eta} = \eta/Y$ where the choice of $Y$ is $Y = g/(2\pi f)^2$ is equivalent to previous studies of McBennett and Harris\cite{McBennettHarris16} and Halev and Harris \cite{HalevHarris18}.  Then \eqref{Eq: ProjectileODEs} becomes
\begin{align}
\ddot{\hat{\xi}} &= -\nu^*\dot{\hat{\xi}};\qquad \hat{\xi}(0) = \hat{x}_n,\quad \dot{\hat{\xi}}(0) = \hat{v}_n^yx\\
\ddot\hat{\eta} &= -G - \nu^*\dot{\hat{\eta}};\qquad \hat{\eta}(0) = \hat{y}_n,\quad \dot{\hat{\eta}}(0) = \hat{v}_n^y,
\end{align}
with
\begin{equation}
\Psi = A^*\cos(2\pi\hat{T})\cos(2\pi\hat{\xi}) + \gamma^*\cos(2\pi[2\hat{T}-\varphi^*]),
\end{equation}
where $\nu^* = 2\nu/f$, $\varphi^* = f\varphi$, $G = 16\pi^2$, $A^* = A(2\pi f)^2/g$, and $\gamma^* = \gamma/g$ are the dimensionless drag coefficient, phase, gravity, wave amplitude, and bath forcing amplitude, respectively.  We observe that in our system $\nu^* \approx 0.1$, and as we will justify in the sequel, the size of $\nu^*$ will exacerbate pathological numerical artifacts that are avoided by considering an undamped model.  We write the dimensionless system with the hats and asterisks removed as
\begin{subequations}
\begin{align}
\ddot{\xi} &= 0;\qquad \xi(0)=x_n,\quad \dot{\xi}(0)=v_n^x\\
\ddot{\eta} &= -G;\qquad \eta(0) = y_n,\quad \dot{\eta}(0) = v_n^y,\\
\Psi &= A\cos(2\pi T)\cos(2\pi\xi) + \gamma\cos(2\pi[2T-\varphi]).
\end{align}
\label{Eq: ProjectileODEsDimensionless}
\end{subequations}

Solving \eqref{Eq: ProjectileODEsDimensionless} yields
\begin{subequations}
\begin{align}
\xi &= v_n^xt + x_n,\\
\eta &= -\frac{1}{2}Gt^2 + v_n^yt + y_n,
\end{align}
\label{Eq: ProjectilePath}
\end{subequations}
which explicitly determines the path of flight of the particle (Fig. \hyperref[Fig: Path]{\ref{Fig: Paths and Waves}a} red dashed line); thereby mitigating the computational expense of solving a system of Ordinary Differential Equations (ODEs).

\begin{figure}[htbp]
\centering
\includegraphics[width = 0.9\textwidth]{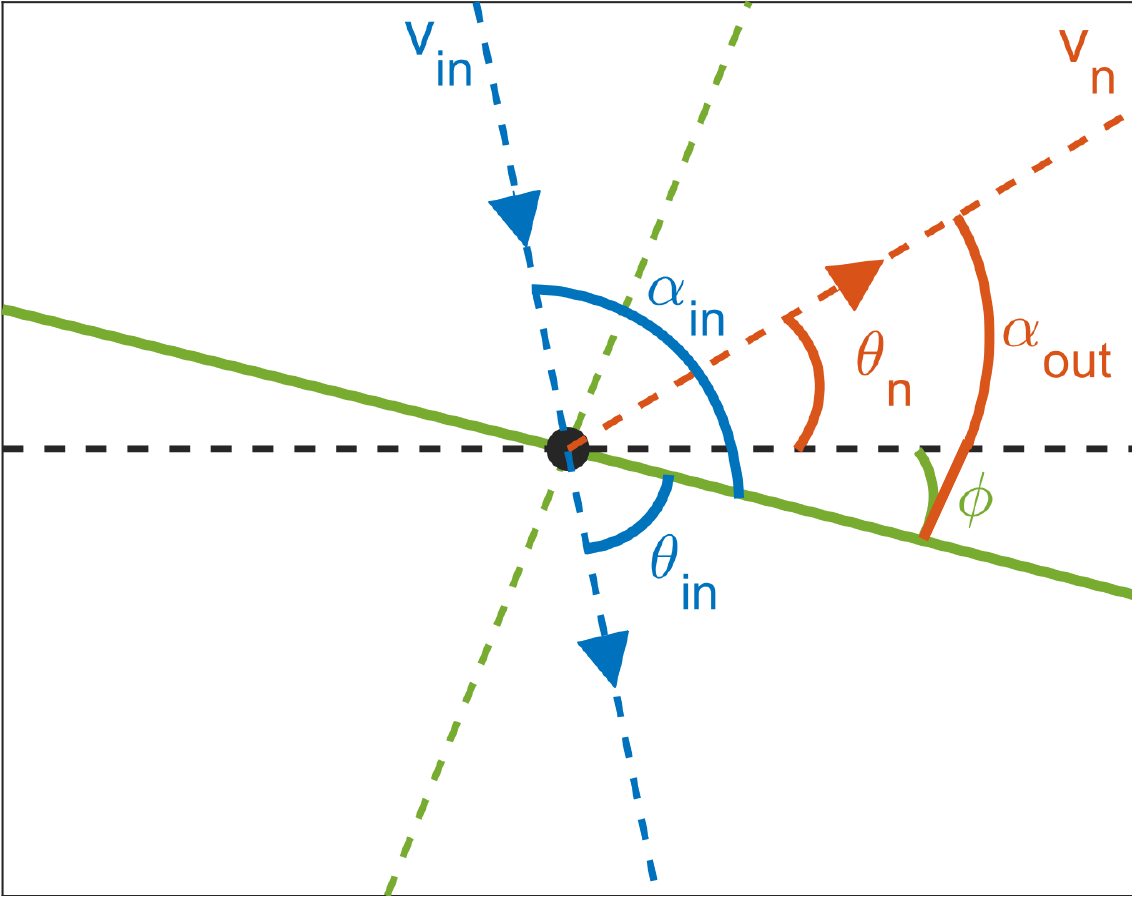}
\caption{The magnified region from Fig. \ref{Fig: Path} with the linearized wavefield (solid green line) and linearized approach path, ${\color{blue}v_{\text{in}}}$, (dashed blue line with direction arrows) intersecting at the impact point (solid black circle).  The abscissa is represented by the horizontal dashed black line, and the normal line of the wavefield at impact is represented by the dashed green line.  Next, the launch direction, ${\color{red}v_n}$ is represented by the red dashed line with a direction arrow.  The right hand acute angle between the linearized wavefield and abscissa is ${\color{green}\phi}$.  The approach angle between the linearized path and abscissa is ${\color{blue}\theta_{\text{in}}}$.  This results in the positive right hand angle, ${\color{blue}\alpha_{\text{in}}} = (\pi + {\color{blue}\theta}) - {\color{green}\phi}$, between the linearized path and linearized wavefield.  Then the launch angle between the launch direction and linearized wavefield, ${\color{red}\alpha_{\text{out}}}$ is calculated.  Finally, the launch angle in cartesian coordinates is given by ${\color{red}\theta_n = \alpha_{\text{out}}} + {\color{green}\phi}$.}\label{Fig: Angles}
\end{figure}

Now, given a particular point of intersection between the path of the projectile and wavefield, we can model the inelastic collision between the droplet and the bath.  As an illustrative aid we employ Fig. \ref{Fig: Angles}.  Suppose the droplet is already in flight with a velocity of $v_{\text{in}}$, just before the $n^\th$ impact,  approaching the point of impact at a slope of $d\eta/d\xi = (d\eta/dt)/(d\xi/dt) = \dot{\eta}/\dot{\xi}$ yielding an approach angle (with respect to the abscissa) of
\begin{equation}
\theta_{\text{in}} = \tan^{-1}\left(\dot{\eta}/\dot{\xi}\right)  \in [-\pi/2, \pi/2].
\end{equation}
furthermore, assume that at the point of impact the wavefield has a tangent line with a slope of $d\Psi/dx$, and hence an angle
\begin{equation}
\phi = \tan^{-1}\left(\frac{d\Psi}{dx}\right) \in [-\pi/2,\pi/2],
\end{equation}
with respect to the abscissa.  The angle between the tangent lines of the path and wavefield is
\begin{equation}
\alpha_{\text{in}} = (\pi + \theta_{\text{in}}) - \phi.
\end{equation}
This gives us the components of the approach velocity $v_{\text{in}}$ that are tangential ($v_{\text{in}}\cos(\alpha_{\text{in}})$) and normal ($v_{\text{in}}\sin(\alpha_{\text{in}})$) to the wavefield.  In Sec. \ref{Sec: Numerics}, we use the velocity components to compute the tangential and normal components of the coefficient of restitution, $C_R^T$ and $C_R^N$.  Then the tangential, which has the opposite sign to that of the approach, and normal components of the launch velocity are
\begin{subequations}
\begin{align}
v_n^T &= -C_R^Tv_{\text{in}}\cos(\alpha_{\text{in}}),\quad \text{and}\\
v_n^N &= C_R^Nv_{\text{in}}\sin(\alpha_{\text{in}}),
\end{align}
\end{subequations}
yielding a launch angle of
\begin{equation}
\alpha_{\text{out}} = \tan^{-1}\left(\frac{v_n^N}{v_n^T}\right) \in [0,\pi],\quad\text{and}\quad
\theta_n = \alpha_{\text{out}} + \phi,
\end{equation}
with respect to the gradient of the wavefield for $\alpha_{\text{out}}$ and the abscissa for $\theta_n$.  Finally, we arrive at the launch velocity components,
\begin{subequations}
\begin{align}
v_n^x &= \sqrt{\left(v_n^T\right)^2 + \left(v_n^N\right)^2}\cos\theta_n,\\
v_n^y &= \sqrt{\left(v_n^T\right)^2 + \left(v_n^N\right)^2}\sin\theta_n,
\end{align}
\end{subequations}
required for \eqref{Eq: ProjectileODEsDimensionless}.

\section{Numerical techniques}\label{Sec: Numerics}

Since the ODEs in \eqref{Eq: ProjectileODEsDimensionless} have explicit solutions, our numerical methods are dedicated to tracking the evolution of the wavefield with respect to the path of the projectile, resolving the effects of each impact, and calculating various statistics to compare with experiments.  The event-driven nature of the phenomenon \cite{MGNB15, GMV2019, FEBMC10}, and inherent discrepancies in sampling between experimental data and numerics, have necessitated novel computational remedies to oft-overlooked simulation problems.  We approach the computational issues from a physics-inspired point of view with the aim of preserving as much of the kinematic realism as possible.

We first need to have a method to find the point of impact as the path continues indefinitely as shown in Fig. \ref{Fig: Path}.  To do this we need to calculate the point of intersection between the flight path and the wavefield, which would be easy if the wavefield stood still.  What makes it even more challenging is the evolution of the wavefield shape in addition to the vertical oscillations.  So, the impact involves three actions: the flight of the particle, the vertical motion of the table, and the evolution of the shape of the wavefield.  Now, we have the quintessential ``not knowing where the edge is until passing it'' problem, and off the edge we go by letting the particle go through the bath.  As soon as the particle height is less than the bath height, $\eta(t_{i+1}) < \Psi(t_{i+1},\xi(t_{i+1}))$, at the horizontal location of the particle, $\xi(t_{i+1})$, the code is paused.  Then it is necessary to work backwards to find the intersection; however, since so many aspects of the system are changing around that impact point, going one time step back to $t_i$ would cause the particle, $(\xi(t_i),\eta(t_i))$, to be too far from the point of intersection.  One remedy is to use a finer global time step, but this would prohibitively increase the computational expense, and instead we choose to use a bisection scheme in the domain $(t_i,t_{i+1})$.  The scheme produces the time $t_*$ such that $\eta(t_*) - \Psi(t_*,\xi(t_*)) < 10^{-4}$, which implies $\eta(t_*) > \Psi(t_*,\xi(t_*))$; a necessary condition to keep the simulation physically realistic.  Finally, the launch variables can be calculated as shown in Sec. \ref{Sec: Kinematics}.

The issue of chattering arises when the velocity of the bath is asymptotically equivalent to the vertical velocity of the droplet at impact; that is, $d\Psi/dt \sim d\eta/dt$ at $t = t_*$.  Then the droplet can numerically go through the bath more than once within one time step, no matter how small we make the time step, which creates a non-isolated intersection; i.e., there is some $t_{**} \neq t_*$ such that $t_i \leq t_{*}\leq t_{**}\leq t_{i+1} = t_i + \varepsilon$ for all $\varepsilon$ larger than some computationally prohibitive time step, and often occurs for all $\varepsilon$ larger than machine accuracy.  That is, numerically, the algorithm records repeated intersections when no such intersection is present physically, which causes the algorithm to effectively halt.  Fortunately, the physical observations offer guidance.  In the laboratory, the impacts are not instantaneous, and therefore a close encounter with the bath may not behave like an impact in our computational setup.  Numerically, we let the droplet shadow the bath for a small period of time until the droplet velocity and bath velocity diverge.  Furthermore, in the previous paragraph we calculated our impact with respect to the center of the droplet.  However, in reality the contact is not with the center, but rather the surface of the droplet, and hence we do the same calculations except with $\Psi + R$ in place of $\Psi$ where $R$ is the radius of the droplet.  The actual physics is more nuanced as the radial separation should be calculated in the direction of the flight path and we do not consider deformations, however this would add considerable numerical complexity with minimal additional insight about the longtime statistics.  Now, if the newly calculated vertical velocity (just after impact) is less than the bath velocity, $v_n^y < d\Psi/dt\vert_{t = t_*}$, which implies that there will be a non-isolated numerical intersection, we let the velocity and position shadow that of the bath, $v_n^y = d\Psi/dt\vert_{t = t_*}$ and $y_n = \Psi(t_*,\xi(t_*)) + R$.

\begin{figure}[htbp]
\centering
\stackinset{r}{1mm}{t}{2mm}{\textbf{ (a)}}{\includegraphics[height = 0.16\textheight]{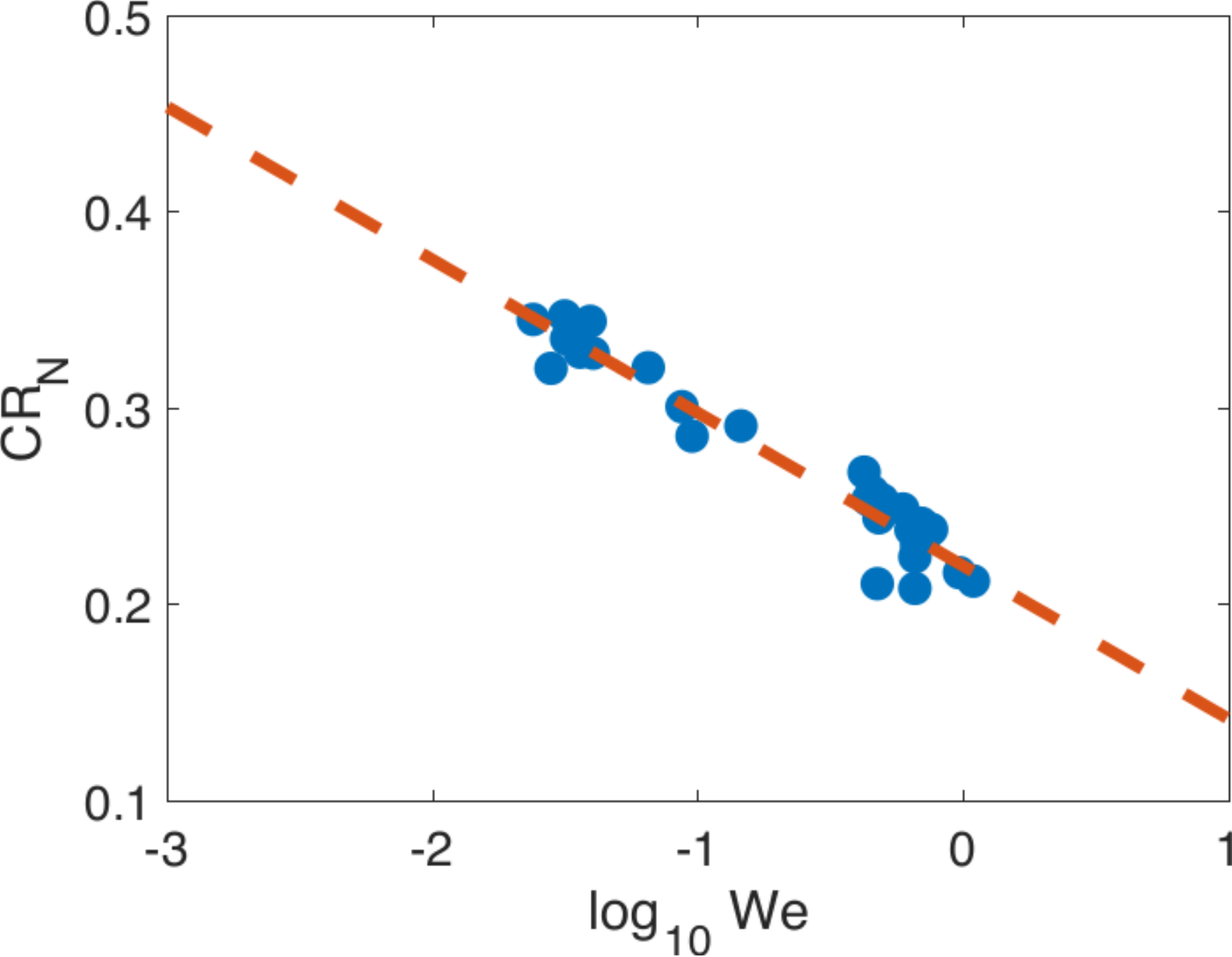}}
\refstepcounter{subfigure}\label{Fig: CR_N}\quad
\stackinset{l}{5mm}{b}{5mm}{\textbf{ (b)}}{\includegraphics[height = 0.16\textheight]{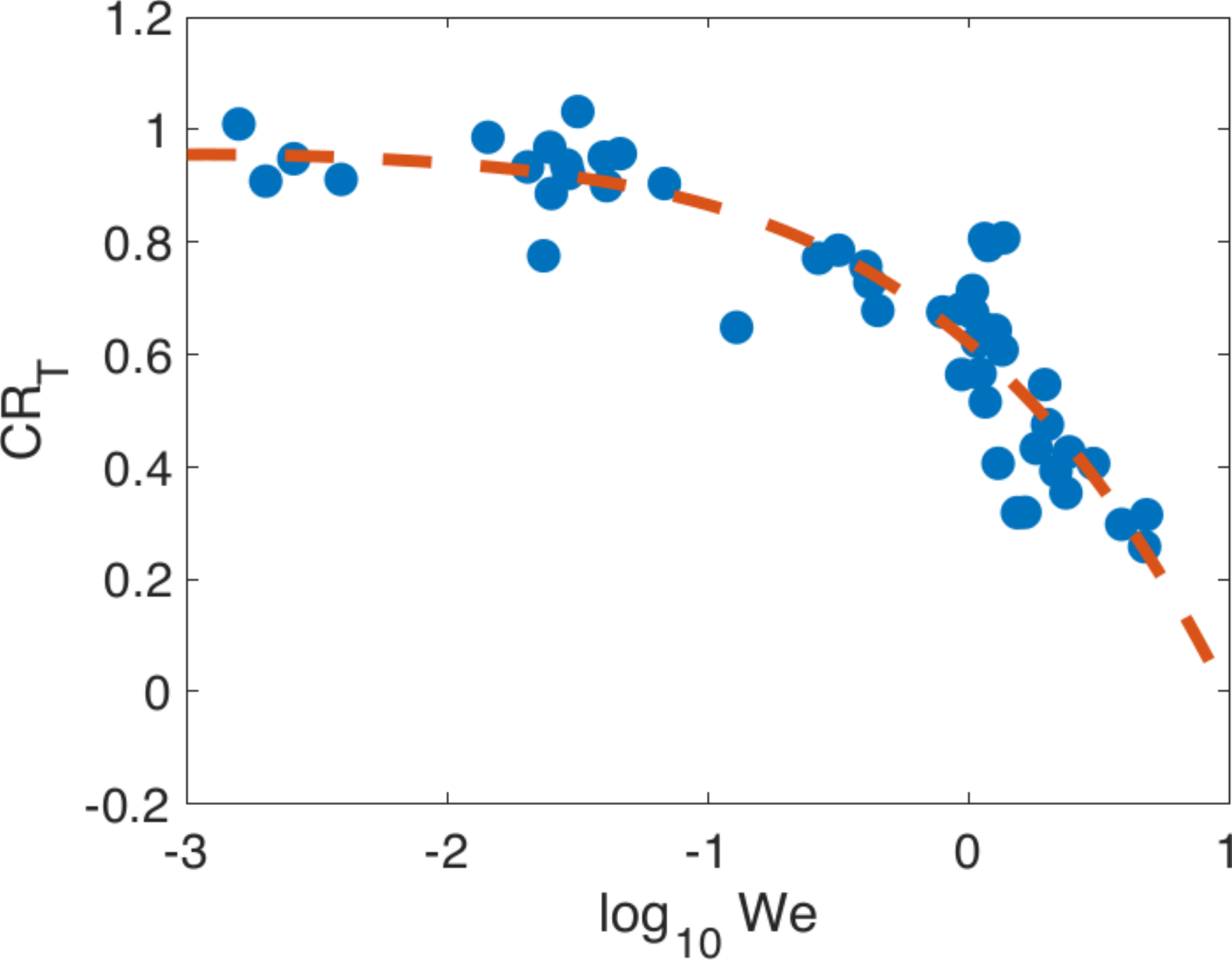}}
\refstepcounter{subfigure}\label{Fig: CR_T}\quad
\stackinset{l}{10mm}{t}{2mm}{\textbf{ (c)}}{\includegraphics[height = 0.16\textheight]{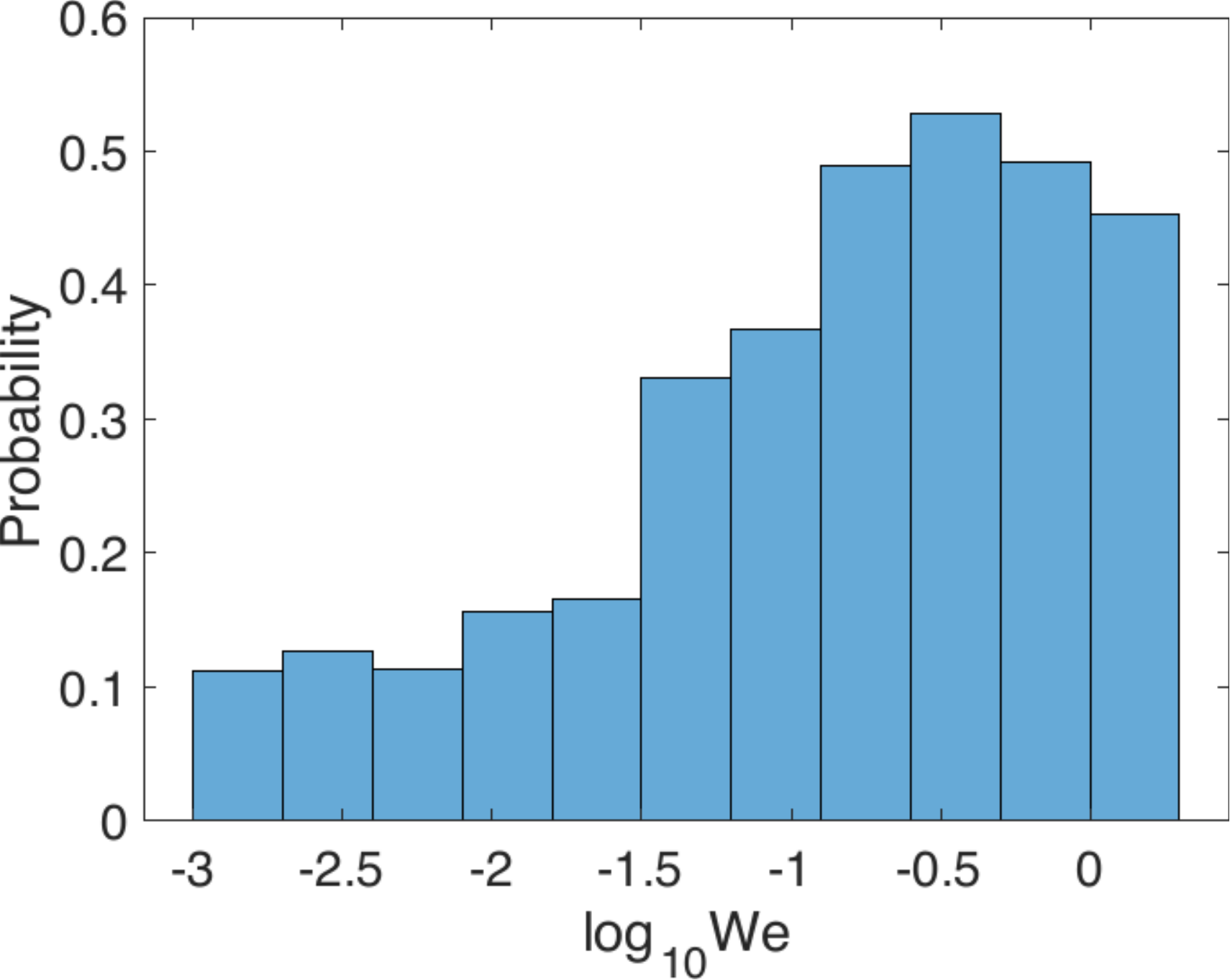}}
\refstepcounter{subfigure}\label{Fig: Webber_hist}
\caption{Normal and tangential coefficients of restitution as a function of $\log_{10}We$.  In figures \textbf{(a)} and \textbf{(b)} blue dots represent the data points from the experiments of Mol\'{a}\v{c}ek and Bush \cite{MolBush13a, MolBush13b}.  \textbf{(a)}  Linear fit (red dashed line) of the data points (blue) for the normal coefficient of restitution.  \textbf{(b)}  Quartic (red dashed curve) fit of the data points (blue) for the tangential coefficient of restitution.  \textbf{(c)}  Histogram of Weber numbers.}\label{Fig: Coefficients of Restitution}
\end{figure}

Next, due to the inelastic nature of the collision between the droplet and the bath, we calculate the coefficients of restitution, $C_R^T$ and $C_R^N$, as functions of the Weber number, $We = \rho R_d \left(v_{\text{in}}^N\right)^2/\sigma$, where $\sigma$ is the surface tension of the droplet and $v_{\text{in}}^N$ is the droplet velocity in the normal direction with respect to the wave surface.  In the experiments of Mol\'{a}\v{c}ek and Bush \cite{MolBush13a, MolBush13b}, the experimental coefficients of restitution were plotted against the Weber number.  We fit various polynomials, with respect to the log of the Weber number, to the data points (Fig. \ref{Fig: Coefficients of Restitution}) and choose a linear fit for $C_R^N$ and a quartic fit for $C_R^T$,
\begin{subequations}
\begin{align}
C_R^N \approx -0.0779\log_{10}We  &+  0.2198,\\
\begin{split}
C_R^T \approx -0.0044(\log_{10}We)^4 &- 0.0454(\log_{10}We)^3\\
&- 0.1906(\log_{10}We)^2\\
&- 0.3948\log_{10}We + 0.6209.
\end{split}
\end{align}
\end{subequations}
Since the polynomial approximations do not asymptote, and it is shown that $0.1 \lesssim We \lesssim 1$ \cite{MolBush13a, MolBush13b} (and also in Fig. \ref{Fig: Coefficients of Restitution}(c)), we restrict the ranges to $0.2 \leq C_R^N \leq 0.34$ and $0.26 \leq C_R^T \leq 0.8$.  Interestingly, the motion of the particle depends more on the tangential coefficient of restitution than on the normal coefficient.

\section{Theoretical results}\label{Sec: Theoretical results}

In this section we use the parameter values listed in Table \ref{Table: Parameters} for the simulations unless otherwise stated.  We first present qualitative similarities between the simulations and experiments.  Then we show statistical results to illustrate the amount of the dynamics that is captured by the model.  Finally, we use the model to discuss subtleties in the dynamics that are inaccessible to experiments.
\begin{table}[htbp]
\centering
\caption{List of frequently used parameters}\label{Table: Parameters}
\begin{tabular}{|lcr|}
\hline
Parameter & Symbol & Value\\
Gravity & $g$ & $9810$ $\text{mm}/\text{s}^2$\\
Bath vibration frequency & $f$ & $80$ $\text{Hz}$\\
Oil density & $\rho$ & $9.5\times 10^{-7}$ $\text{Kg}/\text{mm}^3$\\
Droplet radius & $R$ & $0.28$ $\text{mm}$\\
Droplet surface tension & $\sigma$ & $0.0206$ $\text{Kg}/\text{s}^2$\\
Bath wavelength & $\lambda_f$ & $4.75$ $\text{mm}$\\
Air dynamic viscosity & $\mu$ & $1.84\times 10^{-8}$ $\text{Kg}/(\text{mm}\cdot \text{s})$\\
Bath vibration phase & $\varphi$ & $\pi/2$\\
Dimensionless bath & \multirow{2}{*}{\shortstack{$\varphi^*$}} & \multirow{2}{*}{\shortstack{$40\pi$}}\\ vibration phase & & \\
Effective Gravity & $G$ & $16\pi^2$\\
Dimensionless & \multirow{2}{*}{\shortstack{$A^*$}} & \multirow{2}{*}{\shortstack{$[0, 13)$}}\\ wave amplitude & & \\
Dimensionless & \multirow{2}{*}{\shortstack{$\gamma^*$}} & \multirow{2}{*}{\shortstack{$[4.29, 4.5)$}}\\ bath acceleration & & \\
\hline
\end{tabular}
\end{table}

In addition to the fundamental parameters in Table \ref{Table: Parameters}, we also must resolve the wave amplitude, $A$, which depends on the amplitude of forcing acceleration, $\gamma$.  Since our maximum acceleration is $\gamma = 1.047\gamma_f$, it is convenient to write the coupling factor \cite{WWPMK01} as $\Gamma \approx 1/2\sqrt{1.047 - 1}$, which allows the model to capture the qualitative features observed in experiments.

\begin{figure}[htbp]
\centering
\includegraphics[width=0.95\textwidth]{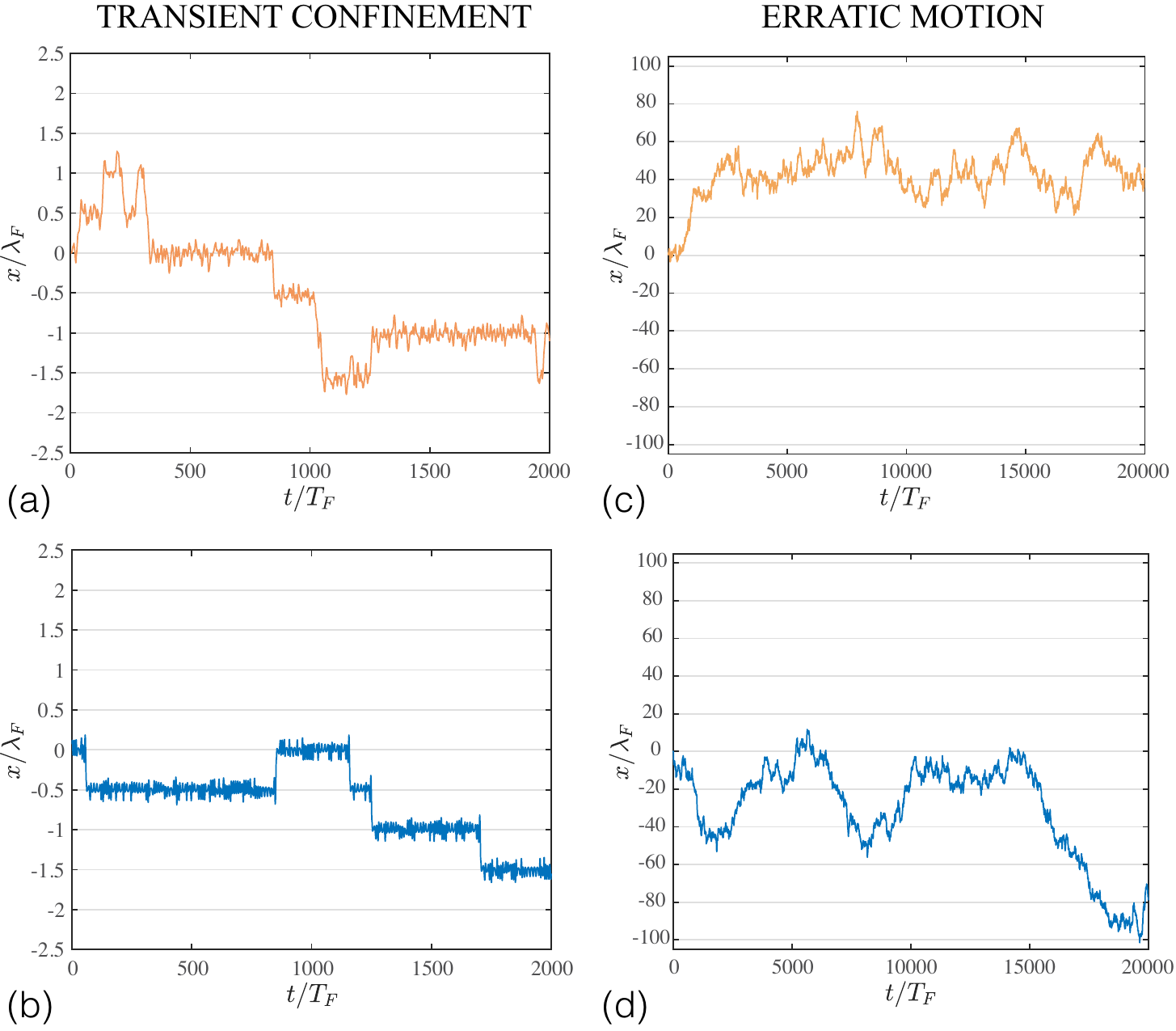}
\caption{The two dynamical regimes of a droplet bouncing on a quasi-1D standing wave: transient confinement (a, b) and erratic motion (c, d). Experimental data: (a) $\gamma/\gamma_F = 1.005$, (b) $\gamma/\gamma_F = 1.047$. Numerical results:  $\gamma/\gamma_F = 1.0055$ (c),  $\gamma/\gamma_F = 1.047$ (d).  See Supplemental Videos for both experiments and numerical simulations of the droplet motion in the erratic regime.}
\label{fig2}
\end{figure}

In both experiments and numerical simulations we identify two regimes of droplet motion, which depend on the dimensionless acceleration $\gamma/\gamma_F$ and thus on the wave amplitude $A$ (Fig. \ref{fig2}). At low amplitude, the droplet executes jumps with length $\sim \lambda_F$ between sites spaced by multiples of $0.5 \lambda_F$. The droplet lingers around each site for a typical time $T_{\rm trap} \gg 10 T_F$ within a region with width $< 0.5 \lambda_F$. Each site is presumably a wave trough within which the droplet is transiently confined in a regime of chaotic bouncing. We call this regime ``transient confinement'' (Fig. \ref{fig2}(a,b)), which is reminiscent of L\'{e}vy flights \cite{MandelbrotLevyFlight}. At higher amplitude the droplet motion is erratic and each site is visited during a typical time $T_{\rm err} \lesssim 10 T_F$ (Fig. \ref{fig2}(c,d)), which is reminiscent of deterministic diffusion \cite{DeterministicDiffusionReview}. 
%---FIGURE2---%
%---%

\begin{figure}[htbp]
\begin{subfigure}{0.4\textwidth}
\stackinset{l}{2mm}{t}{1mm}{\textbf{ (a)}}{\includegraphics[height = 0.16\textheight]{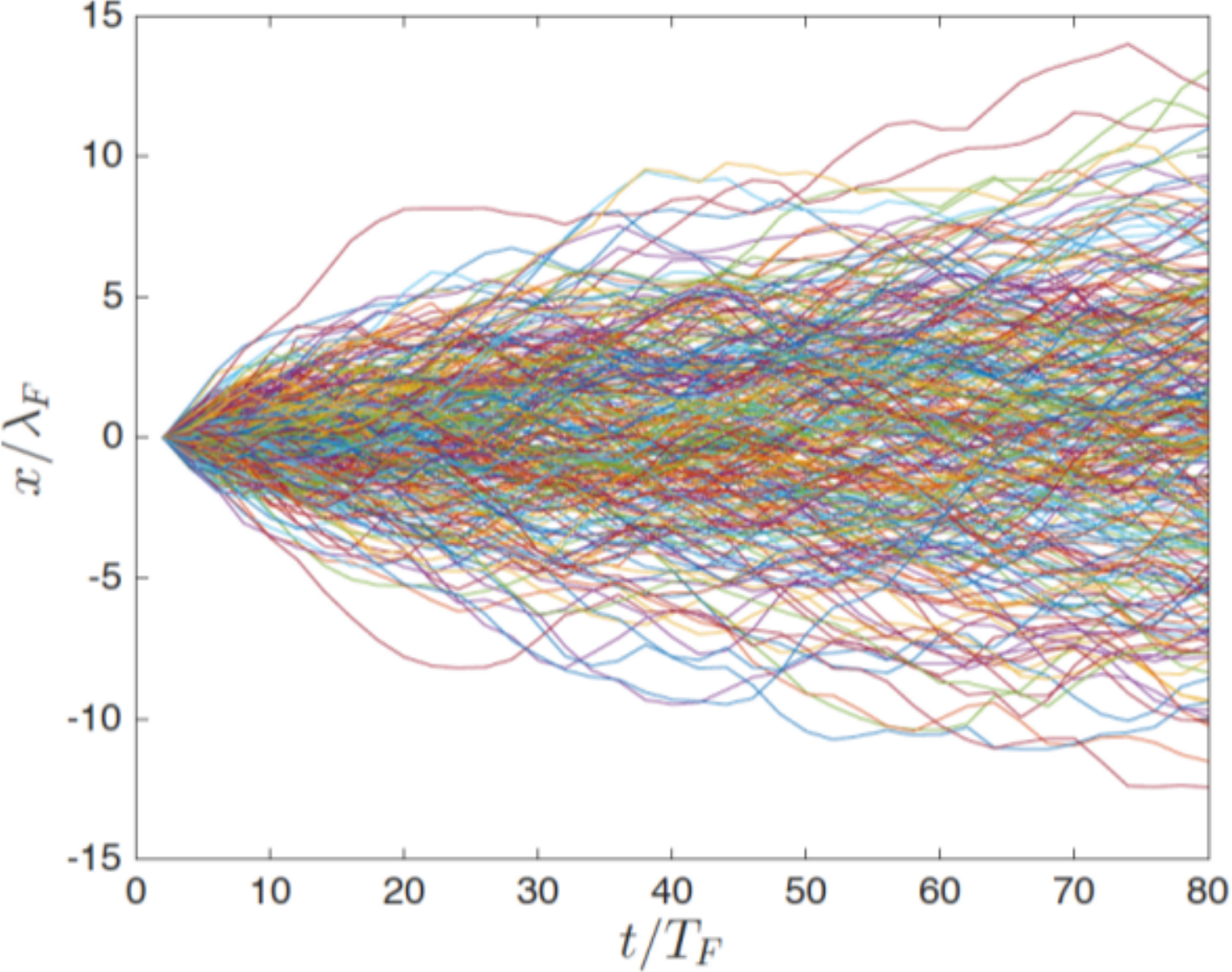}}

\stackinset{l}{3mm}{t}{1mm}{\textbf{ (b)}}{\includegraphics[height = 0.16\textheight]{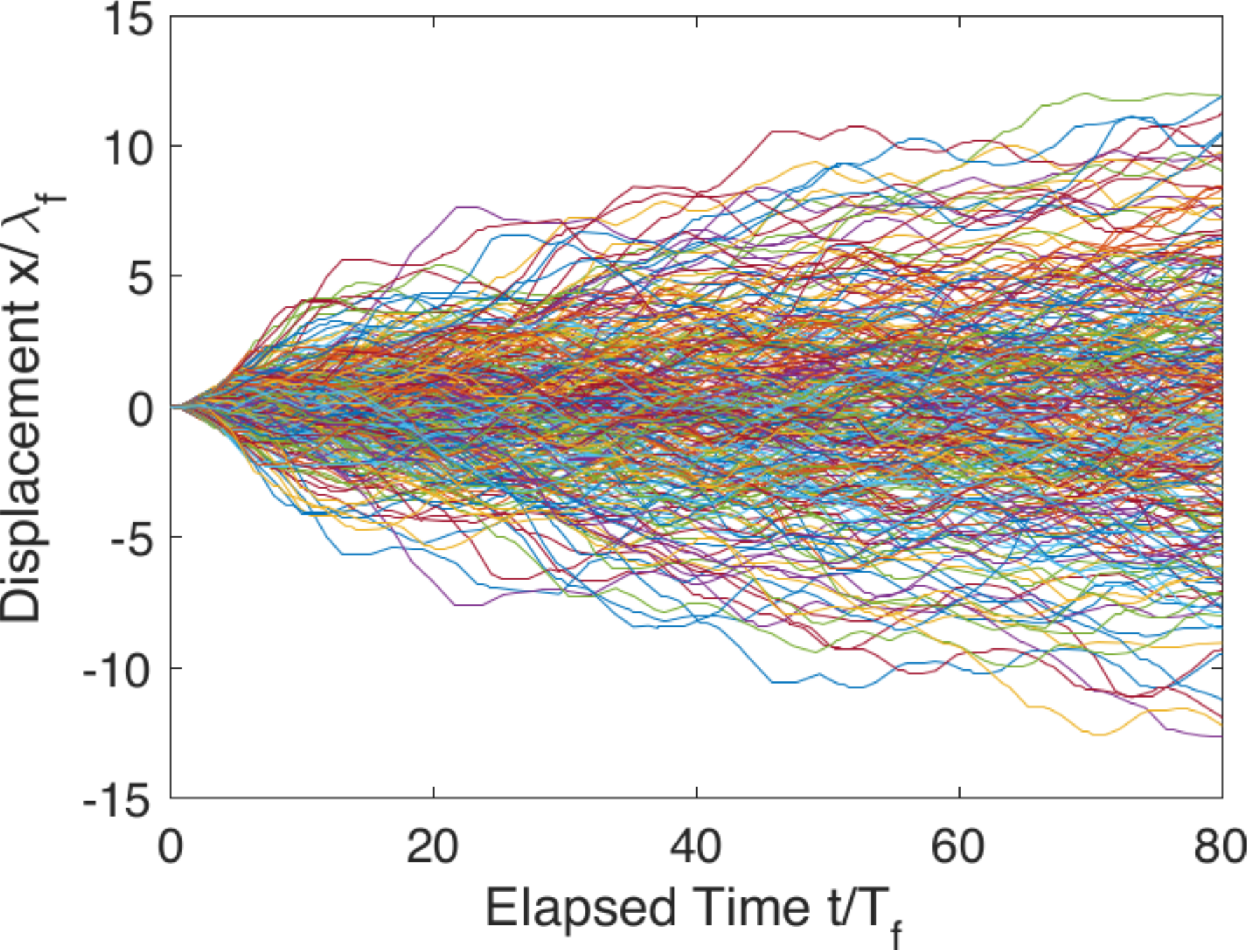}}
\end{subfigure}\hspace{-18pt}
\begin{subfigure}{0.5\textwidth}
\stackinset{l}{8mm}{t}{2mm}{\textbf{ (c)}}{\includegraphics[height = 0.32\textheight]{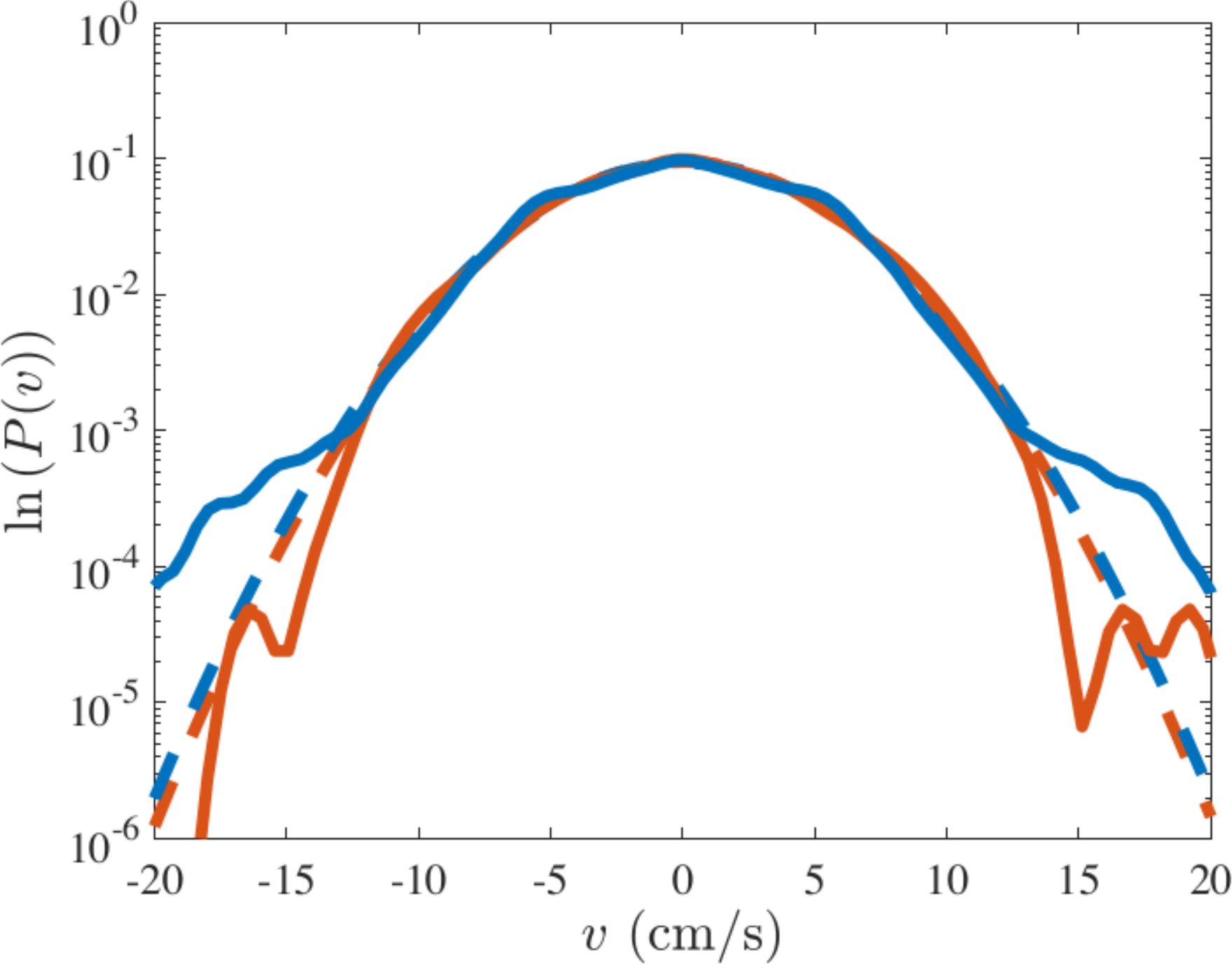}}
\end{subfigure}

\begin{subfigure}{0.9\textwidth}
\stackinset{l}{15mm}{t}{2mm}{\textbf{ (d)}}{\includegraphics[width = \textwidth]{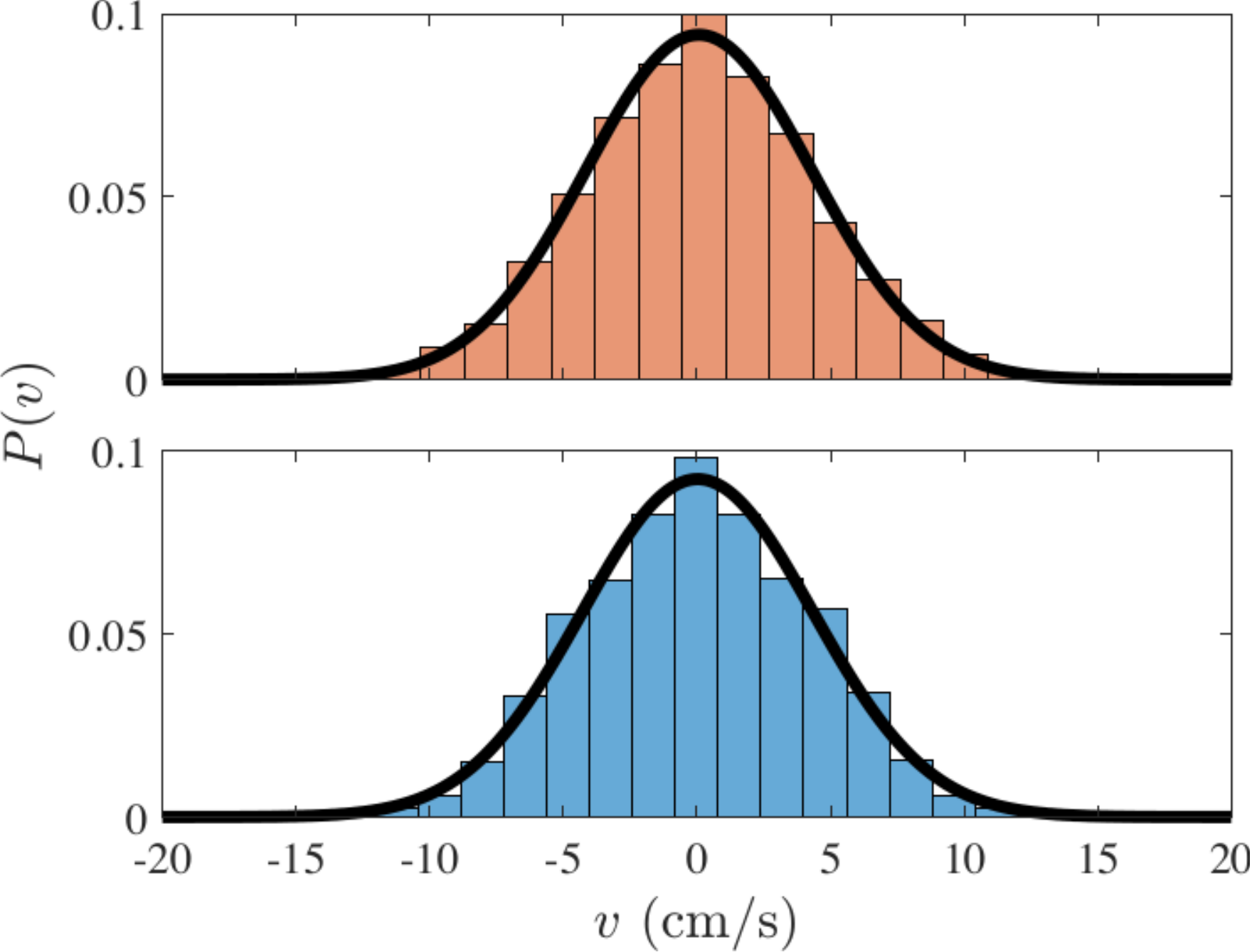}}
\end{subfigure}
\caption{Evidence of diffusive behavior from long-time trajectories in experiments and simulations.  \textbf{(a)}  Dispersion of particles for different experimental runs and \textbf{(b)} different initial points in simulations.  \textbf{(c)}  Log of the probability density function for experiments (solid red curve) and simulations (solid blue curve) plotted on top of their respective fits (dashed curves).  \textbf{(d)} Histograms of the velocities for $\gamma = 1.047\gamma_f$.  The experimental distribution in red has a mean of $0.0725942$cm/s with a standard deviation of $4.23167$cm/s.  The simulations have a mean of $0.028891$ cm/s with a standard deviation of $4.31803$cm/s.  The velocities are recorded from $300$ experimental runs/simulations, the trajectories of which are then allowed to flow for $T_{\max} = 80$ (i.e., $80$ table oscillation periods) with average velocity in experiments calculated between video frames at a frame rate of $20 Hz$, and $1/30$s intervals in simulations to avoid effects from the table forcing frequency.}\label{Fig: Histograms and Dispersion}
\end{figure}

The walker appears to prefer velocities near zero and chooses higher speeds with decreasing probability as illustrated in Fig. \ref{Fig: Histograms and Dispersion}(a, b), where the trajectory from Fig. \ref{fig2}(c, d) is one of the trajectories from Fig. \ref{Fig: Histograms and Dispersion}(a, b).  This is very similar to particles experiencing Brownian motion.  Moreover, Fig. \ref{Fig: Histograms and Dispersion}(c, d) shows that the distribution of the velocities for $\gamma = 1.047\gamma_f$ appear to be Gaussian, which is indicative of Brownian motion.

\begin{figure}[htbp]
\centering
\includegraphics[width=0.95\textwidth]{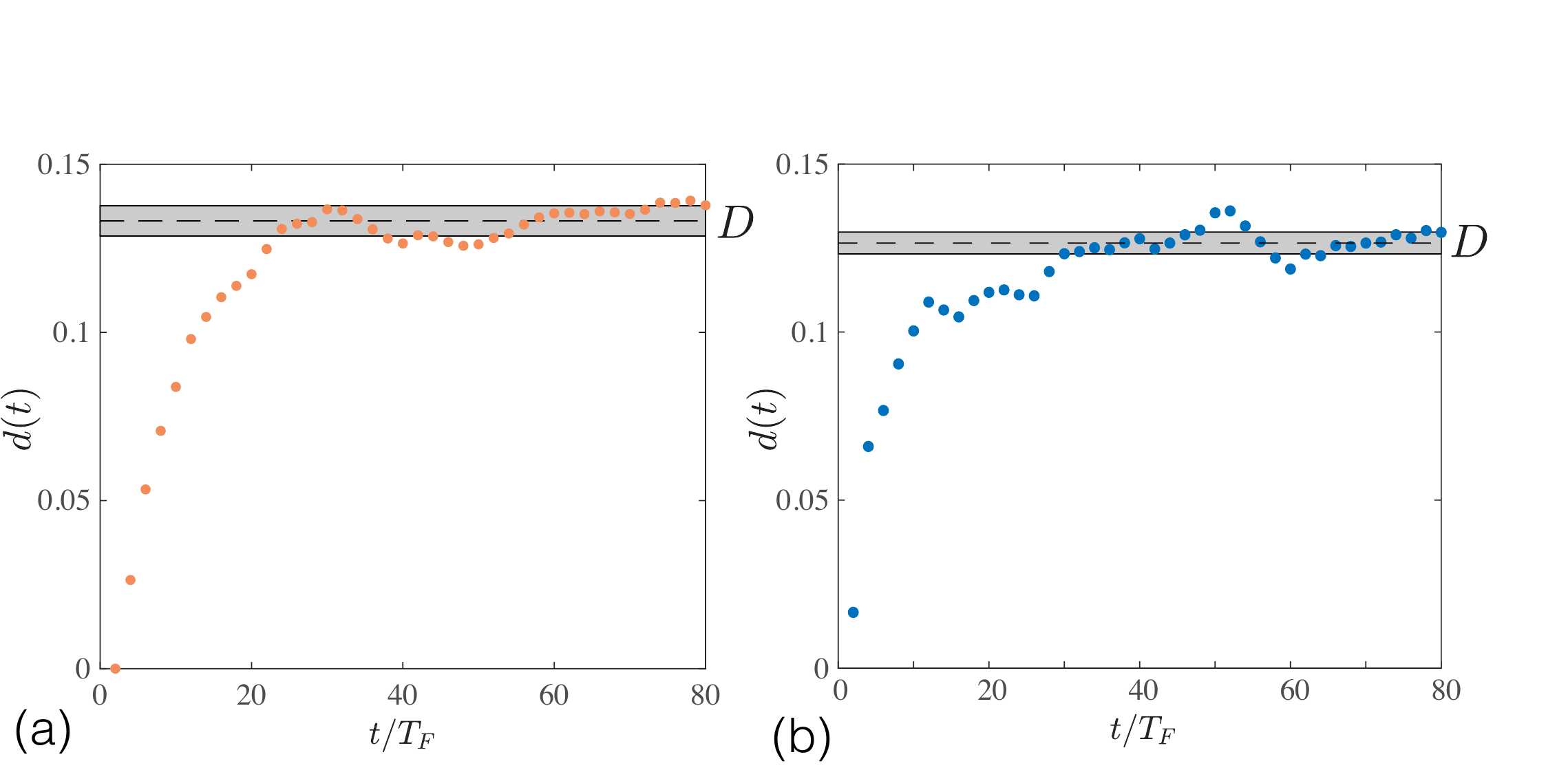}
\caption{Temporal evolution of the quantity $d(t)$ in \eqref{Eq:  Diffusivity}. Experimental data (a) and numerical result (b) with $\gamma/\gamma_F=1.047$. The dashed line is the mean of $d(t)$ over the last 20 points, which corresponds to the diffusion coefficient $D$, and the grey area covers one standard deviation from the mean.}
\label{fig3}
\end{figure}

In order to quantify the behavior of the system in terms of diffusion we compute the quantity
\begin{equation}
    d(t) = \frac{T_F}{\lambda_F^2}\frac{\sigma^2(t)}{2t},
    \label{Eq:  Diffusivity}
\end{equation}
where $\sigma^2(t)=\sum_{i=1}^N \vert x_i - \bar{x} \vert^2 / (N-1)$, with $N$ the number of points and $\bar{x}$ the mean position. Experimentally, this is measured by splitting a video with duration $\geq60$~s in samples with duration $2$~s and computing the variance across the samples. We observe that $d(t)$ increases with time to saturate around a constant mean value (Fig. \ref{fig3}(a)). 
Using the dimensionless model in \eqref{Eq: ProjectilePath}, $d(t)$ is computed numerically by first applying the model to $300$ equally-spaced initial points within an interval corresponding to one wavelength, $x_0 \in [0,1)$, for an elapsed time corresponding to $80$ bath oscillations, $T = 80$. We then compute the variance $\sigma^2$ for each run and divide by $2T$ to obtain $d(t)$ as in \eqref{Eq:  Diffusivity}.
%---FIGURE3---%
%---%
The temporal evolution and saturation value of $d(t)$ in experiments and numerical simulations are similar. We call 
\begin{equation}
    D=\lim_{t \to \infty} d(t)    
\end{equation}
the diffusion coefficient, that we compute in both experiments and numerical simulations as the average value of $d(t)$ over the last $20$ time steps.

\begin{figure}[htbp]
\centering
\includegraphics[width=0.95\textwidth]{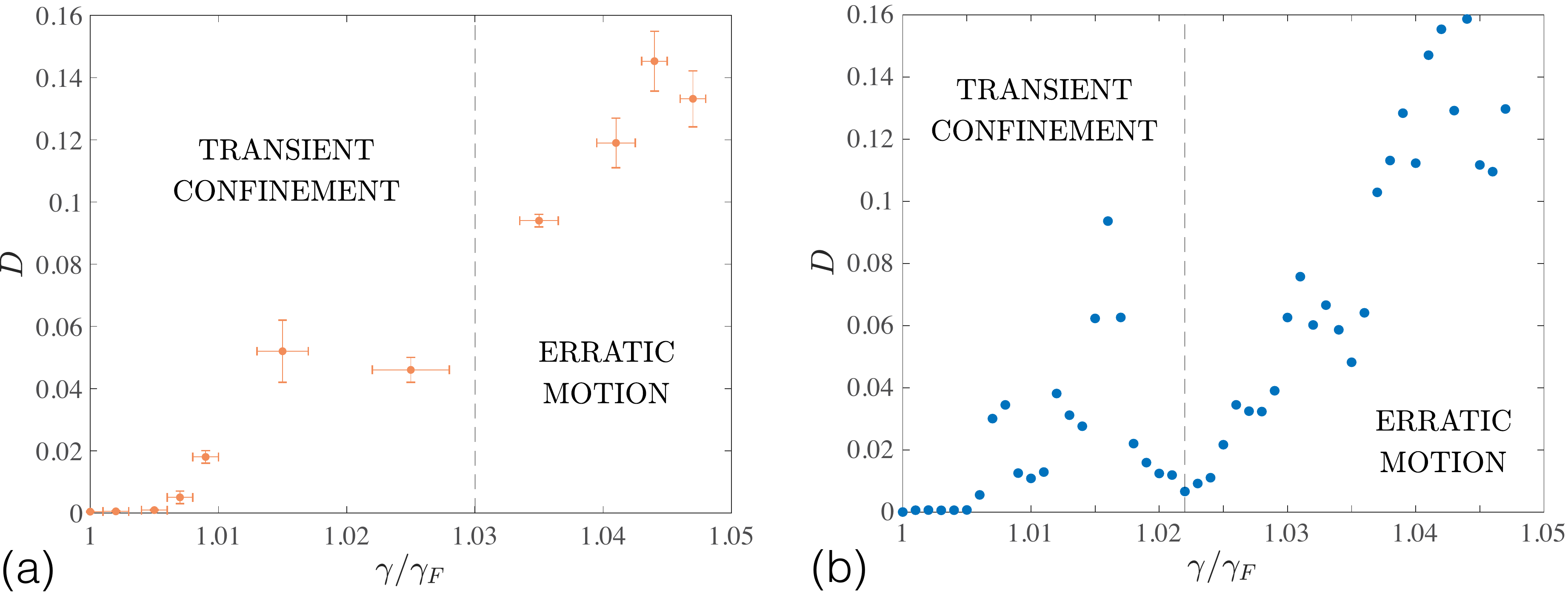}
\caption{Diffusion coefficient $D$ as a function of the dimensionless forcing acceleration $\gamma/\gamma_F$. (a) Experimental data. (b) Numerical results.}
\label{fig4}
\end{figure}

We proceed by varying the forcing acceleration $\gamma/\gamma_F$ in both experiments and simulations and computing $D(\gamma/\gamma_F)$. The results are presented in Fig. \ref{fig4}. In both experiments and numerical simulations we observe an overall increase of the diffusion coefficient with the forcing acceleration, in both the transient confinement and the erratic motion regimes. As an exception to this general trend, we observe $D$ to decrease at the edge between the two regimes. In experiments, this is due to long-lasting confinement in which the droplet presumably bounces only once over four bath vibration periods. In the model when the particle starts moving, its velocity is nearly constant yielding an appreciable $\sigma(t)$, however when the motion starts to become erratic (at the edge between the two regimes), the droplet velocity has equal probability of being positive and negative, effectively reducing $\sigma(t)$.  As $\gamma/\gamma_F$ is increased furthermore, the droplet velocity is large enough that once again the deviation from most initial points is appreciable, which contributes to an increasing $\sigma(t)$ even though there will still be some initial points that do not contribute.
%---FIGURE---%

\section{Conclusion}\label{Sec: Conclusion}

We often find chaotic dynamics, induced by simple interactions, giving rise to quite complex phenomena.  This is particularly prevalent in deterministically diffusive systems as described in a review by Geisel \cite{DeterministicDiffusionReview} and by Artuso and Cvitanovi\'{c} \cite{ChaosBookDeterministicDiffusion}.  This is also true for a larger class of damped-driven systems \cite{LiWaiKutz10, RJorB18, koch2020mode}, which can be studied using an energy gain-loss formulation.

In this work we studied the motion of a droplet bouncing on a standing wave. As the wave amplitude is increased the motion of the droplet is observed to go through several bifurcations until it eventually becomes erratic.  At this erratic regime the distribution of droplet velocities tends to a Gaussian, which indicates potentially diffusive behavior.  In fact, the effective diffusivity can be computed using \eqref{Eq:  Diffusivity}, and we observed that the diffusivity is not a monotonic function of the bath forcing.  We then presented a detailed derivation of the hydrodynamic-kinematic model of walking droplets on an annular cavity with bath forcing above the Faraday wave threshold.  Then we went over the numerical methods used to simulate the circumferential droplet motion leading to long-time statistics.  The numerical methods require unique solutions to mitigate the chattering effect due to the event-driven nature of the problem.  These mitigation strategies were inspired by the physical process being investigated thereby preserving much of the physical agreement with experiments.  The theoretical results are studied in more detail in Sec. \ref{Sec: Theoretical results}.  We showed that there are three dynamical regimes in the model:  trapped (minor excitation around a fixed velocity), transient confinement (Levy-like flights \cite{MandelbrotLevyFlight} where the droplet is trapped for variable amounts of time), and the diffusive state (droplet motion is seemingly random with chaotically changing velocity).  Notice that the trapped state is just a special case of the transient confinement state.  We show that the qualitative behavior of the model is very close to that of experiments. This modeling framework can also be used for fast computation of long-time statistical behavior when the vertical motion of the particle is necessary, which is something a stroboscopic model may overlook.

Interestingly, our hydrodynamic-kinematic model exhibits properties, such as Brownian-like motion and Levy flights, similar to those of other deterministically induced diffusive systems \cite{DeterministicDiffusionReview}.  In addition to future dynamical systems studies, the present experimental and modeling framework can be used to conduct a more detailed 2-dimensional investigation of droplets above the Faraday wave threshold.  While Tambasco \textit{et al.} \cite{TambascoDiffusion} also observed diffusive behavior above the Faraday wave threshold, they use a much larger droplet ($20 \%$ larger at the lower end).  By using a smaller droplet we are able to ignore deformation and drag during the modeling process.  Furthermore, we foresee a more detailed investigation of diffusive properties for the motion of the droplet than done here yielding far more nuanced and unexpected dynamics.

Finally, there are also intriguing pedagogical possibilities.  Both the experiment and model can be reproduced to be used in a teaching lab as a visual demonstration of a system exhibiting diffusion.  The experimental setup can be reproduced using an inexpensive audio speaker and a wave generator on a mobile phone.  Using 3-dimensional printing, a dish with an annular cavity to hold the bath can be produced.  The dish would then be attached to the speaker (either by glue or a 3-D printed adapter).  When the waves are sent through the mobile phone to the speaker the bath will be accelerated vertically creating the environment for walking droplets.  Moreover, the full hydrodynamic-kinematic model of Sec. \ref{Sec: Kinematics} can be easily simulated with the available computationally efficient MATLAB codes.  To study a variety of parameters one can modify the easy to follow preamble of the code before running it as usual.  Most importantly, the model is accessible to a wide range of individuals, and we hope that it will be particularly useful as a pedagogical example of a damped-driven system for undergraduate students and even high school students.

\section*{Acknowledgment}

First and foremost, this work could not have been accomplished without the experimental data from Giuseppe Pucci, to whom A.R. is very grateful.  A.R. also thanks John Bush for the use of the facilities in MIT’s Applied Math Lab.  In addition, this work has benefited from fruitful discussions with Giuseppe Pucci, Dan Harris, and J. Nathan Kutz.  Finally, A.R. appreciates the support of the Amath department at UW.

\bibliographystyle{unsrt}
\bibliography{Energy_gain-loss}

\end{document}